\documentstyle[prb,aps,psfig]{revtex}

\begin{document}
\wideabs{
\title{Phase-transitions induced by easy-plane anisotropy
       in the classical Heisenberg antiferromagnet on a
       triangular lattice: a Monte Carlo simulation}

\author{Luca Capriotti\cite{e-LC}}
\address{Dipartimento di Fisica dell'Universit\`a di Firenze
         and Istituto Nazionale di Fisica della Materia (INFM),\\
         Largo E. Fermi~2, I-50125 Firenze, Italy,\\
         and Scuola Internazionale Superiore di Studi Avanzati,
         via Beirut 2-4, 34013 Trieste, Italy}

\author{Ruggero Vaia\cite{e-RV}}
\address{Istituto di Elettronica Quantistica
         del Consiglio Nazionale delle Ricerche,
         via Panciatichi~56/30, I-50127 Firenze, Italy,\\
         and Istituto Nazionale di Fisica della Materia (INFM).}

\author{Alessandro Cuccoli\cite{e-AC}, Valerio Tognetti\cite{e-VT}}
\address{Dipartimento di Fisica dell'Universit\`a di Firenze
         and Istituto Nazionale di Fisica della Materia (INFM),\\
         Largo E. Fermi~2, I-50125 Firenze, Italy}

\date{\today}
\maketitle

\begin{abstract}
We present the results of Monte Carlo simulations for the antiferromagnetic
classical XXZ model with easy-plane exchange anisotropy on the triangular
lattice, which causes frustration of the spin alignment. The behaviour of
this system is similar to that of the antiferromagnetic XY model on the
same lattice, showing the signature of a Berezinskii-Kosterlitz-Thouless
transition, associated to vortex-antivortex unbinding, and of an Ising-like
one due to the chirality, the latter occurring at a slightly higher
temperature. Data for internal energy, specific heat, magnetic
susceptibility, correlation length, and some properties associated with the
chirality are reported in a broad temperature range, for lattice sizes
ranging from $24{\times24}$ to $120{\times}120$; four values of the
easy-plane anisotropy are considered. Moving from the strongest towards the
weakest anisotropy (1\,\%) the thermodynamic quantities tend to the
isotropic model behaviour, and the two transition temperatures decrease by
about 25\,\% and 22\,\%, respectively.
\end{abstract}


} 


\section{Introduction}
\label{s.intro}

The critical behaviour of classical two-dimensional (2D) frustrated models
has raised the interest of several scientists in the last years. Popular
realization of such models are Heisenberg
\cite{KawamuraM84,AzariaDM92,WintelEA95} and the XY
\cite{MiyashitaS84,LeeJNL8486} antiferromagnet on a triangular lattice, as
well as the fully frustrated XY model on the square lattice
\cite{TeitelJ83,Olsson9597}; the role of frustration shows up in the
particular nature of the order parameter in the first one, and in the
presence of two kinds of symmetry in the other two. In the
triangular-lattice models the minimum energy configuration (say, the ground
state) is a ferromagnetic arrangement of the spins in each of three
sublattices, with a relative rotation of 120$^\circ$ between each other.
The three vertices of each lattice plaquette belong to different
sublattices, and it is possible to associate to each plaquette a vector,
the chirality, that defines the rotation of the spin direction around the
plaquette. In the XY case the (staggered) chirality turns into a scalar
order parameter whose sign distinguishes between two degenerate ground
states.

In this paper we consider the XXZ triangular antiferromagnet (TAF), a
system that shares the symmetry of the XY model and is a more realistic
description of a spin system. In addition, to treat three-dimensional spins
is a necessary step for studying the corresponding quantum system by
means of the pure-quantum self-consistent harmonic approximation
\cite{CGTVV95,BCTVV96,CTVV96prl,CCTVV97}.
In the XXZ TAF the easy-plane anisotropy results in a double degeneracy of
the ground state. This model is thus expected to belong to the universality
class of the frustrated XY model. A very fascinating aspect of the
thermodynamics of such a system is that it has two order parameters, the
(in-plane) magnetization and the chirality, with two distinct symmetry
groups, continuous $SO(2)$ rotations and discrete $Z_2$ lattice
reflections, respectively. This has suprising consequences in view of
Mermin-Wagner's theorem \cite{MerminW66}, that can be applied only in the
first case to infer that the magnetization must vanish at any nonzero
temperature, according to the expected Berezinskii-Kosterlitz-Thouless
(BKT) \cite{BKTall} critical behaviour associated with the rotation
symmetry in the $xy$ plane, while long-range order and an Ising-like phase
transition are allowed for the chirality. This rich structure was observed
in Monte Carlo (MC) simulations for the XY model. However, it was not clear
whether the two transitions are distinct, with an intermediate phase, or
they are manifestations of a new universality class, in which the two
transition temperatures could be molten in a single multicritical point:
from early numerical simulations \cite{MiyashitaS84,LeeJNL8486} they turned
out to be weakly different, but in view of their uncertainty no firm
conclusions could be inferred. Very recently, high-precision MC studies of
the XY TAF model and of its Villain version were reported by Olsson
\cite{Olsson9597}, who established that the BKT transition occurs at a
temperature $\sim$\,1--1.4\,\% lower than the Ising-like one.

Qualitatively, the results we find for the XXZ TAF are rather similar to
those already known for the XY TAF, for any anisotropy strength considered,
even close to the isotropic limit. The observed changes reflect the
enhanced spin fluctuations out from the easy plane and the crossover
towards the isotropic limit. The features of both the Ising-like and the
BKT transition shift to lower temperatures. Although the numerical
uncertainty is not much smaller than their difference, from our simulations
the BKT critical temperature turns out slightly smaller than the other one,
systematically for all anisotropy values.

In Sec.~\ref{s.xxzmodel} we introduce the XXZ model and its ground state,
discussing the connections with the corresponding XY model and its critical
behaviour. In Sec.~\ref{s.MC} the MC simulation algorithm is described and
the definitions of the calculated thermodynamic quantities are given.
Eventually, in Sec.~\ref{s.rescomm} the MC results are reported, generally
for four different values of the anisotropy constant, and analyzed for
their critical behaviour and the finite-size effects. Conclusions are
briefly drawn in Sec.~\ref{s.concl}.


\section{The XXZ Model}
\label{s.xxzmodel}

\subsection{Definition of the model}
\label{ss.definimod}

We consider the classical XXZ Hamiltonian
\begin{equation}
 {\cal H}={J\over2}\sum_{{\bf{i}},{\bf{d}}}
 \big( s_{\bf{i}}^x s_{{\bf{i}}+{\bf{d}}}^x
 + s_{\bf{i}}^y s_{{\bf{i}}+{\bf{d}}}^y
 +\lambda \, s_{\bf{i}}^z s_{{\bf{i}}+{\bf{d}}}^z \big)~,
\label{e.xxzmodel}
\end{equation}
where $J$ is the positive (antiferomagnetic) exchange constant, and
$\big(s_{\bf{i}}^x,s_{\bf{i}}^y,s_{\bf{i}}^z\big)$ are the Cartesian
components of unitary vectors, the classical spins, sitting on the
sites $\{{{\bf{i}}}\}$ of a two-dimensional triangular lattice.
The interaction is restricted to nearest-neighbours and ${\bf{d}}$ runs
over their relative displacements ($|{\bf{d}}|=1$). The use of unitary spin
vectors produces no loss of generality in the study of a classical system
since the case of generic spin $S$ can be taken into account by rescaling
the exchange constant $J\rightarrow{JS^2}$. The exchange constant sets the
natural energy scale of the system: hence, in the following, energies and
temperatures will be always given in units of $J$.

The planar character of the system is due to the presence of the constant
$\lambda\in[0,1)$, which weakens the interaction of the $z$ spin
components, energetically favouring configurations with the spins lying in
the $xy$ plane (easy-plane). For $\lambda=1$ the isotropic Heisenberg model
\cite{KawamuraM84,WintelEA95} is recovered; for $\lambda=0$ (when the model
is also known as XX0) the spin components on the $z$ axis do not even
appear in the Hamiltonian, making it formally identical to that of the XY
model, which was the subject of some previous works
\cite{MiyashitaS84,LeeJNL8486}. However, in the XY and XX0 models the
phase-spaces are different: in the former the spins are two-dimensional
vectors, while in the latter they can fluctuate out of the $xy$ plane. Of
course the thermodynamic properties are expected to be quantitatively
different in the two models.

\subsection{Ground state properties}
\label{ss.gsprop}

\begin{figure}
\centerline{\psfig{bbllx=5mm,bblly=90mm,bburx=205mm,bbury=188mm,%
figure=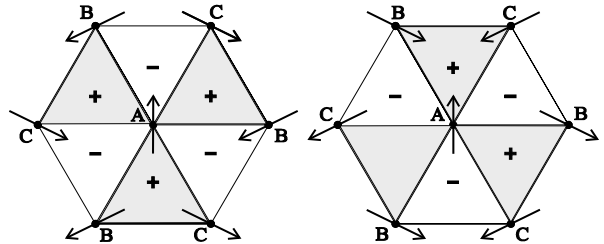,width=80mm,angle=0}}
\caption{
Two degenerate ground states. The plus and minus denote the sign of the
chirality of each elementary cell and the two configurations are
characterized by opposite staggered chirality. The letters A, B and C label
the three sublattices on which the spins are ferromagnetically aligned. }
\label{f.groundstate}
\end{figure}

The ground state configuration of (\ref{e.xxzmodel}) can be found
\cite{Wannier50} by minimizing the energy of any single elementary
triangular cell of the lattice. In this way one gets that, for every value
of $\lambda\in[0,1]$, the ground state consists of coplanar spins forming
$\pm{2}\pi/3$ angles between nearest-neighbours (see
Fig.~\ref{f.groundstate}). In contrast to the isotropic case, where the
plane in which the $2\pi/3$ structure lies can take any direction in spin
space, in the XXZ model (as well as in the XY model) such structure must
take place in the easy-plane. In any case this leads to a
$\sqrt{3}\times\sqrt{3}$ periodic ground state.

In the XY and XXZ unfrustrated systems, like the ferromagnet on a general
lattice or the antiferromagnet on a bipartite lattice, the degeneracy of
the ground state, connected to the possible equivalent choices of the
direction of alignment of spins in the space, corresponds to the $SO(2)$
symmetry group. In the planar TAF the frustration effect causes an
additional two-fold degeneracy of the ground state, which is due to
chirality (or helicity), defined as the sign of rotation of the spins along
the sides of each elementary triangle. Since global spin rotations in the
easy-plane conserve the chirality, one configuration cannot be obtained
from the other one by pure rotations but it is necessary to include some
other symmetry operation such as lattice reflection; i.e. the whole
degeneracy corresponds to the group $SO(2)\times{Z_2}$.

\subsection{Phase transitions}
\label{ss.phtrans}

As it has been pointed out by Lee {\em et al.} \cite{LeeJNL8486}, the
existence of the extra degeneracy of the ground state allows the 2D
frustrated XY model to support, in addition to spin waves and vortices, a
third type of elementary excitation associated with domain-wall formation
between regions with opposite staggered chirality (solitons). The first two
are responsible for the loss of orientational order with increasing
temperature, while the latter causes the destruction of chirality order.
Then, the thermodynamics is characterized by the interplay of these three
types of elementary excitations.

The ground state of the three sublattices consists of ferromagnetically
aligned spins, the interaction between those of the same sublattice being
mediated by the spins of the other two. Thus the physics of the three
sublattices is expected to be similar to the physics of the whole lattice
in the XY unfrustrated models and a Berezinskii-Kosterlitz-Thouless (BKT)
topological transition becomes plausible \cite{BKTall}. Besides, since
lattice reflection is a discrete symmetry, the system can have long range
chirality order at finite temperature without violating the Mermin-Wagner
theorem \cite{MerminW66} and an order-disorder Ising-like transition is
expected. Indeed this is the situation observed for the first time in the
MC study performed by Miyashita and Shiba \cite{MiyashitaS84} for the
36-state clock model and, later, by Lee {\em et al.} \cite{LeeJNL8486}, on
the basis of group-theoretical symmetry arguments, combined with more
extensive MC simulations including somewhat larger lattices (up to
$72\times72$) and without the limitations introduced by the discrete
sampling of the spin orientation. Perhaps the most striking characteristic
of the phase transition is the logarithmic divergence of the specific heat,
quite similar to that of the Ising model, and for this reason it was
associated with the order-disorder transition of the chirality. The sharp
drop with increasing temperature of the staggered chirality, the
counterpart of Ising magnetization, and the divergence of the corresponding
chiral susceptibility are the other main characteristics of the transition
connected with the loss of chirality order. Instead, the main sign of the
vortex-antivortex unbinding driven BKT transition is the divergence of the
spin correlation length and susceptibility \cite{MiyashitaS84,LeeJNL8486}.
Whether the two transitions are very close but distinct, each conserving
the proper critical exponents, or they melt in a single phase-transition
which has both Ising and BKT character in a possible new universality
class, is an open problem also shared by the other realizations of the
so-called fully frustrated XY model, which are generally believed to have
similar critical behaviour \cite{Olsson9597}.

It is well known \cite{PelcovitsN76,Khokhlacev76,HikamiT80,CTVmcxxz} that the
unfrustrated XXZ model shows the BKT transition, as the XY model, at finite
temperature for every value of the anisotropy constant $\lambda<1$ (no
matter how close to 1); in particular, for $\lambda \rightarrow 1$, the
critical temperature is predicted to approach $0$ as $1/|\ln(1-\lambda)|$.
The phase transition in the XXZ model is again due to vortex unbinding as
in the pure XY model, even if the vortices may have different character
since the spins are no longer obliged to lie in the $xy$ plane, with the
spins in the vortex core pointing preferably in the out of plane direction
\cite{Wysin94}.
Likewise, it is at least plausible that the order-disorder transition of
the chirality is also present in the frustrated XXZ model for $\lambda<1$;
i.e., as far as the system shows a planar character and its ground state
(the $2\pi/3$ structure of the ground state being forced to lie in the $xy$
plane) has the two-fold additional degeneracy.


\section{Monte Carlo simulation}
\label{s.MC}

\subsection{Simulation algorithm and procedure}
\label{ss.simalg}

We performed standard MC simulations on triangular lattices of size
$L\times{L}$ (along the primitive vectors directions), containing $N=L^2$
sites and $2N$ elementary cells, with periodic boundary conditions. $L$ was
between 24 and 120: multiples of 3 have been chosen for $L$ to preserve the
ground-state translation symmetry. In order to reduce the MC correlation
time, a combination of Metropolis \cite{Metropolis53} and over-relaxed
algorithm was used \cite{BrownC87,Creutz87}, as in previous studies of the
XXZ model on the square lattice \cite{CTVmcxxz}. For any given lattice size
and for any temperature, the simulation procedure was the following. An
initial configuration was generated at random and then a given number,
typically 10\,000, of Metropolis moves (one move consists of $N$
single-particle moves) were made in order to bring the system to a
thermalized configuration; after that, the accumulation of the averages
started. A sample configuration, used to update the averages of the
thermodynamic quantities, was taken after $N_{\rm{O}}=4$ over-relaxed moves
and $N_{\rm{M}}=2$ Metropolis moves, the values of $N_{\rm{O}}$ and
$N_{\rm{M}}$ being a compromise between computational effort and
minimization of MC correlation time. Typical runs sampled 30\,000
configurations. Since to our knowledge, no data is available for the XXZ
TAF model, the simulation code we developed was checked in the isotropic
case, for which we got results in complete agreement with those reported in
the literature \cite{KawamuraM84,WintelEA95}.

The strategy adopted to choose the lattice sizes and the temperatures
during the simulations was the following. We performed a sequence of
simulations for the smallest lattice size ($L=24$) with a big step in
temperature, in order to get a clue about the qualitative behaviour of the
various quantities and to locate, at least approximatively, the critical
region. Then it has been necessary to perform various simulations in the
neighbourhood of the critical temperature and for larger lattice sizes, in
order to locate the transition temperatures and to evaluate the finite-size
effects which strongly affect the behaviour of some thermodynamic
quantities.

In the following, the symbol $\left<Q\right>$ will denote the MC
average of a general quantity $Q$, defined as
\begin{equation}
\left<Q\right> = \frac{1}{M} \sum_{j=1}^{M} Q_{j}~,
\end{equation}
$M$ being the total number of configurations $j$ sampled during the
simulation. The uncertainties reported in the following are statistical
errors, estimated in the standard way from the quadratic fluctuations of
the corresponding observable. The effects of correlations were included
multiplying the pure statistical errors by $\sqrt{2\tau}$, $\tau$ being the
correlation time deduced from the analysis of the sequence of the sampled
data for that observable \cite{MadrasS88}.

\subsection{Thermodynamic observables}
\label{ss.thermo}

We evaluated several thermodynamic quantities in order to observe the
effects of the destruction of chirality order, such as internal energy,
specific heat, chirality and its susceptibility. On the other hand, spin
correlation functions, correlation length and susceptibility, were
calculated to investigate the presence of a BKT transition. The internal
energy per spin is defined as
\begin{equation}
 e = \frac{\big<E\big>}{N}~,
\end{equation}
where $E={\cal H}/J$.
The specific heat can be computed, using the relation
\begin{equation}
 c=\frac{1}{N} \frac{\big<E^2 \big>
 - \big< E \big>^2}{t^2}~,
\end{equation}
with the reduced temperature $t=T/J$.

The definition of the chirality, that is the sign of rotation of spins on
the elementary triangle in the ground state configuration, is usually
generalized \cite{KawamuraM84,MiyashitaS84} to nonzero temperatures as
follows:
\begin{equation}
{\mbox{\boldmath{$\kappa$}}}_{\bf r} = \frac{2}{3 \sqrt{3}}
\big( {\bf s}_{1} \times {\bf s}_{2} +
{\bf s}_{2} \times {\bf s}_{3} +
{\bf s}_{3} \times {\bf s}_{1}
\big)~,
\label{e.chir}
\end{equation}
where 1,2 and 3 are the corner sites of the elementary cell centered at
the dual-lattice vector ${\bf r}$, always (for both upward and downward
triangles) ordered in the same way (for example, counterclockwise). Unlike
the XY model where, being the spins confined in the $xy$ plane, it has only
the component along the $z$ axis, in the XXZ model the chirality is a true
vector. It is normalized to 1 for a complete $2\pi/3$ structure, when it
has only the $z$ component that can take the values $\pm 1$, as the Ising
spin. At any finite temperature the length of $\kappa^z$ compared to that
of the other two components gives a measure of the rigidity of the $2\pi/3$
structure, and can be taken as the order parameter. During the simulations
we calculated the staggered chirality, defined as
\begin{equation}
 {\kappa} = \frac{1}{2N}\bigg< \bigg| \sum_{{\bf{r}}} (-)^{\bf r}
 {\kappa}_{{\bf{r}}}^{z} \bigg| \bigg>~,
\label{e.staggchir}
\end{equation}
where the factor $(-)^{\bf r}$ assumes the values $\pm{1}$ for
downward/upward triangles, respectively.

Also interesting is the susceptibility associated to staggered chirality
along the $z$ axis, calculated as
\begin{equation}
 \chi_{\bf \kappa} =
 \frac{1}{4N} \bigg[ \bigg< \bigg| \sum_{\bf r} (-)^{\bf r}
 \kappa_{\bf r}^{z} \bigg|^2\, \bigg> -
 \bigg< \bigg| \sum_{\bf r} (-)^{\bf r}\kappa_{\bf r}^{z}
 \bigg| \bigg>^2\, \bigg]~.
\label{e.staggsuscchir}
\end{equation}

On the other hand, the sublattice in-plane spin susceptibility was computed
as the average of the sublattice squared magnetization,
\begin{equation}
 \chi = {1\over2}\sum_{\alpha=x,y}~\frac{1}{N} \sum_{\Lambda = A,B,C}
 \bigg< \bigg( \sum_{{\bf{i}} \in \Lambda}
 s_{{\bf{i}}}^{\alpha} \bigg)^2\, \bigg>~,
\label{e.chimc}
\end{equation}
since the average magnetization in the thermodynamic limit is 0. Such a
definition retains its value in investigating the divergence of $\chi$ also
for finite lattice simulations, because, even if different from 0, the
missing term in Eq.~(\ref{e.chimc}) is well behaved and negligible in the
critical region. Alternatively, the same information can be also obtained
from the total ${\bf{k}}$-dependent susceptibility,
\begin{equation}
 \chi({\bf{k}}) = {1\over2}\sum_{\alpha=x,y}~
 \frac{1}{N} \bigg< \bigg| \sum_{{\bf{i}}} s_{{\bf{i}}}^{\alpha}
 e^{i{\bf k} \cdot {\bf{i}}} \bigg|^2\, \bigg>~,
\end{equation}
taken at the ordering wavevector ${\bf{K}}$, i.e., one of the six vectors
pointing towards the corners of the first Brillouin zone of the whole
lattice; for example ${\bf{K}}=\big(4\pi/3,0\big)$. A straightforward
calculation shows indeed that the sublattice susceptibility, defined in
Eq.~(\ref{e.chimc}), and the total one satisfy the following relation
\begin{equation}
 \chi({\bf K}) = {3\over2}\,\chi - {1\over2}\,\chi({\bf{k}}{=}0)~,
\label{e.relchi}
\end{equation}
where $\chi({\bf{k}}{=}0)$ is 0 at $t=0$, and is small with
respect to the first, also near to the critical point.

The spin correlation length is defined assuming the asymptotic exponential
decay form of the in-plane spin correlation functions,
\begin{equation}
 C({{\bf{n}}})=\left<s_{{\bf{i}}}^{x}s_{{\bf{i}}+{\bf{n}}}^{x}
 + s_{{\bf{i}}}^{y}s_{{\bf{i}}+{\bf{n}}}^{y} \right> \propto
 e^{- n/\xi }~,
\label{e.expdecay}
\end{equation}
with ${\bf{i}}$ and ${\bf{i}}+{\bf{n}}$ belonging to the same sublattice,
and large values of ${\bf{n}}$. Even if a direct fit of the two-point
correlation function (\ref{e.expdecay}) can be used, we adopted a faster
and more reliable method to evaluate $\xi$. This can be achieved
translating Eq.~(\ref{e.expdecay}) in the reciprocal space. In fact, the
asymptotic exponential decay in real space is associated with the
Ornstein-Zernicke form of the Fourier transform of the (full lattice) spin
correlation function, i.e., the ${\bf{k}}$-dependent susceptibility
$\chi({\bf{k}})$, behaving as
\begin{equation}
 \chi({\bf K}{+}{\bf k}) \propto \frac{1}{ k^2 + \xi^{-2} }~,
\label{e.chitrans}
\end{equation}
for small values of the wave vector ${\bf{k}}$. Since the first
Brillouin zone of the finite lattice is discrete, it is not possible to
take arbitrarily small values of ${\bf{k}}$: we used a fit with the first
four shells around ${\bf{k}}=0$. An alternative way we used was to
extract the value of $\xi$ using just the smallest available ${\bf{k}}$:
\begin{equation}
 \xi = \frac{1}{k_1} \bigg[
 {\chi({\bf{K}})\over\chi({\bf{K}}{+}{{\bf k}_{1}})} - 1 \bigg]^{1/2}~,
\label{e.xifast}
\end{equation}
with, e.g., ${\bf{k}}_{1}=(0,4\pi/L\sqrt3)$.


\section{Results and comments}
\label{s.rescomm}

\subsection{Thermodynamic behaviour}
\label{ss.thermobeha}

We performed simulations on the XXZ model for values of $\lambda=0$, 0.5,
0.9, 0.99, ranging from the strongest easy-plane anisotropic case to the
quasi-Heisenberg case. The isotropic model was also considered not only to
compare the results of our simulation code with the data reported in the
literature \cite{KawamuraM84,AzariaDM92,WintelEA95} but also to check the
consistency of the quasi-Heisenberg limit.

\begin{figure}
\centerline{\psfig{bbllx=32mm,bblly=33mm,bburx=175mm,bbury=258mm,%
figure=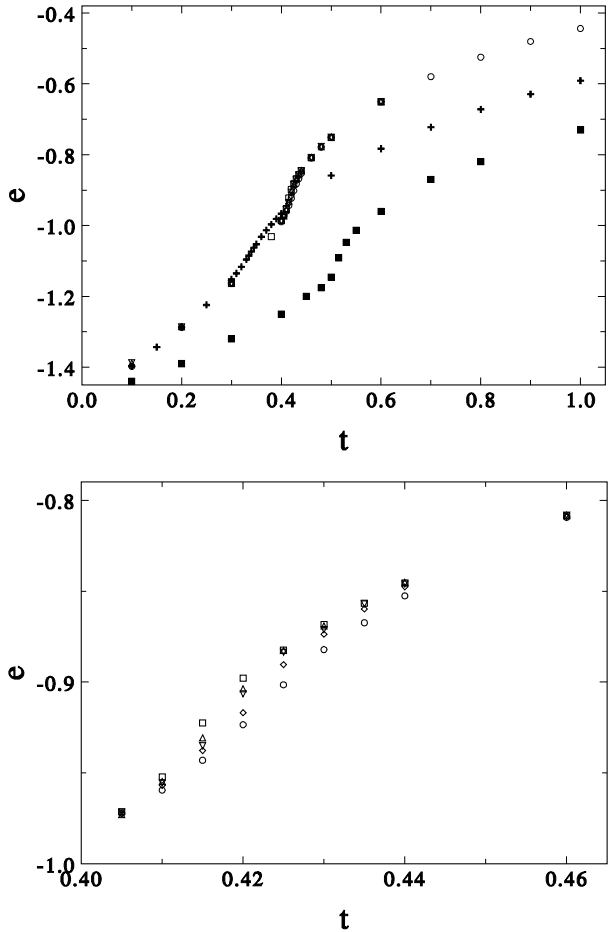,width=80mm,angle=0}}
\caption{
Internal energy for the XX0 model for lattice sizes $L=24$ (circles),
$L=36$ (diamonds), $L=48$ (down triangles), $L=60$ (up triangles) and
$L=120$ (squares). Data for the XY model with $L=30$, taken from
Ref.~\protect\onlinecite{MiyashitaS84} (full squares), and for the isotropic
case with $L=60$ (crosses) are also shown for comparison. In the
lower figure the data are reported in a magnified scale in order to
emphasize the size-dependence. }
\label{f.intenergy}
\end{figure}

The internal energy per spin $e$ for the XXZ model with $\lambda=0$, is
reported in Fig.~\ref{f.intenergy} as a function of the reduced
temperature $t$. For comparison, data for the XY model taken from
Ref.~\onlinecite{MiyashitaS84} and for the isotropic model are also shown.
It is evident from the figure that the qualitative behaviour of the
internal energy is quite different in the isotropic and planar cases. In
fact in the latter cases the internal energy presents a narrow region in
which size-dependence is apparent, and the slope becomes steeper the larger
the lattice size. Instead, the behaviour of the same quantity in the
isotropic case is smoother and only with a weak size dependence (not shown
in the figure). In the low temperature region the internal energy,
according to the spin wave approximation and the equipartition theorem, is
linear in $t$. While in the XY model the internal energy starts from the
ground state value $e_{0}=-3/2$ with slope $1/2$, in the other two cases
the slope is $1$, because of the different number of degrees of freedom:
one per spin in the former, two in the latter cases. Increasing the
temperature, the excitation of the out-of-plane component of the spins
becomes more and more important and causes the different behaviour between
the isotropic and XX0 case. Similar behaviour is observed for all the
values of $\lambda \neq 0$ considered.

\begin{figure}
\centerline{\psfig{bbllx=56mm,bblly=36mm,bburx=151mm,bbury=262mm,%
figure=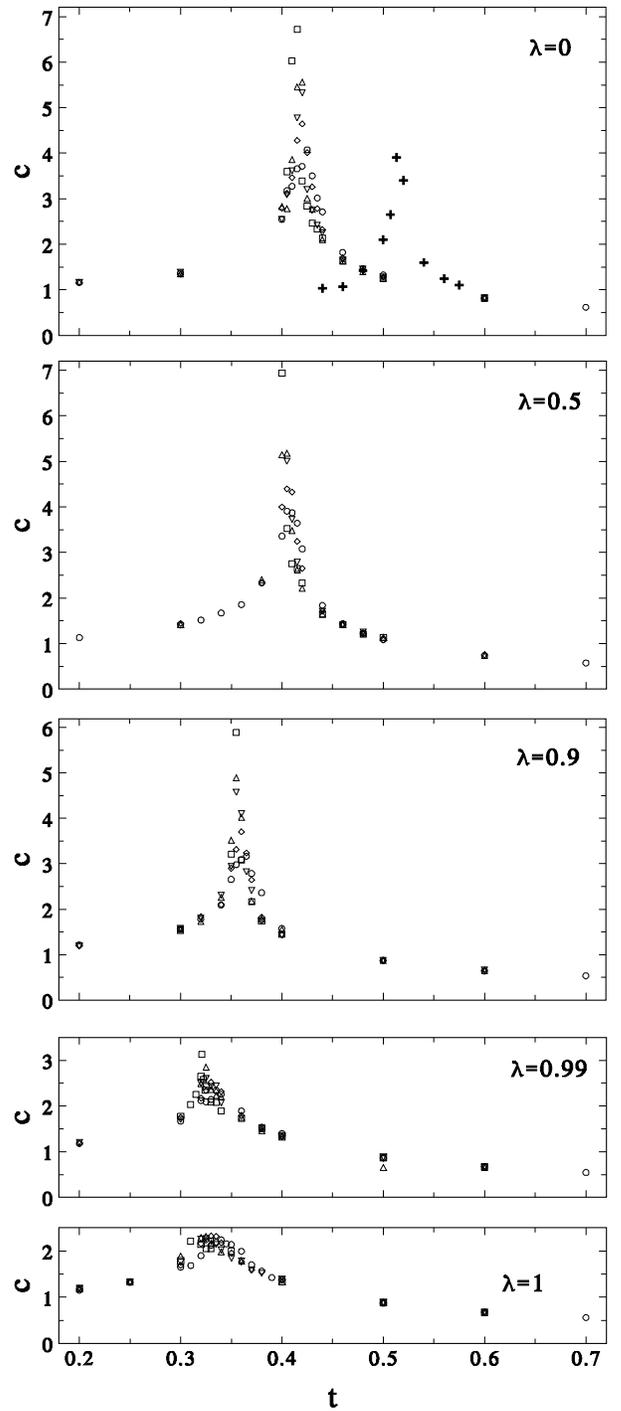,width=80mm,angle=0}}
\caption{
Specific heat for the XXZ model for the reported values of $\lambda$ and
different values of $L$. In the top graph the specific heat observed in the
XY model, for $L=72$, taken from Ref.~\protect\onlinecite{LeeJNL8486}, is
also shown (crosses). Open symbols as in Fig.~\protect\ref{f.intenergy}. }
\label{f.spheat}
\end{figure}

\begin{figure}
\centerline{\psfig{bbllx=16mm,bblly=70mm,bburx=193mm,bbury=204mm,%
figure=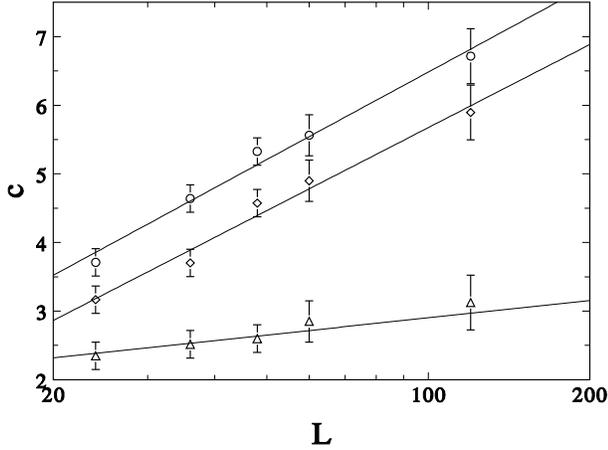,width=80mm,angle=0}}
\caption{
Maximum of the specific heat as a function of $L$ for $\lambda=0$
(circles), 0.9 (diamonds) and 0.99 (triangles). The straight lines are
guides for the eye. }
\label{f.maxsphsc}
\end{figure}

The specific heat data are reported in Fig.~\ref{f.spheat} for all the
values of $\lambda$ (including the isotropic case). For $\lambda \neq 1$
the specific heat shows the signature of a divergence which is an important
feature of the frustrated planar antiferromagnet, also present in the XY
case (the corresponding peak, taken from Ref.~\onlinecite{LeeJNL8486}, is
shown in the figure on the top). As $\lambda \rightarrow 1$ the peak and
the size dependence of its height become less and less pronounced, until,
in the isotropic limit, no divergence at all is observed.
The size dependence of the peak height is shown in Fig.~\ref{f.maxsphsc},
for $\lambda=0$, 0.9 and 0.99. It suggests a logarithmic divergence with
$L$, just as in the two-dimensional Ising model \cite{Barber83}.

\begin{figure}
\centerline{\psfig{bbllx=22mm,bblly=20mm,bburx=185mm,bbury=261mm,%
figure=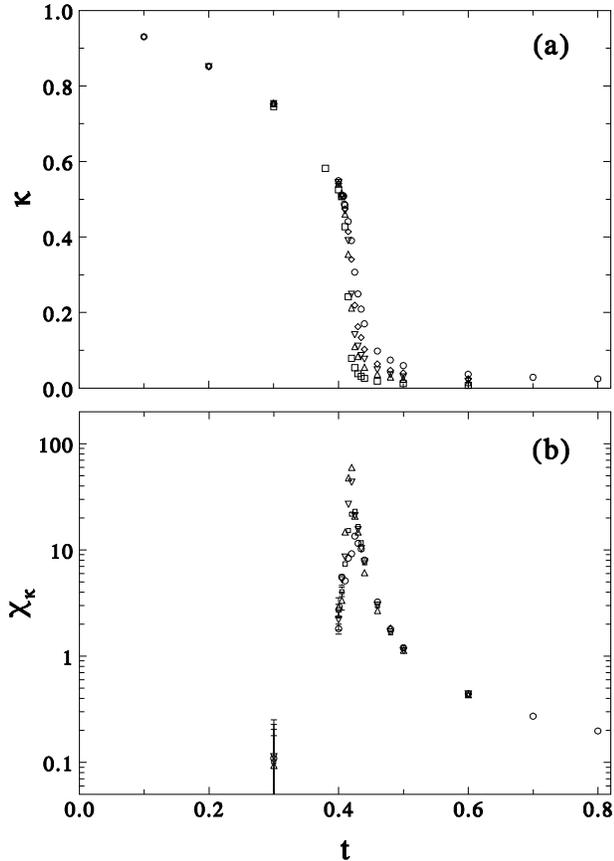,width=80mm,angle=0}}
\caption{
Staggered chirality (a) and chiral susceptibility (b) for $\lambda=0$ and
various values of $L$, as functions of temperature. Symbols as in
Fig.~\protect\ref{f.intenergy}. }
\label{f.chiral}
\end{figure}

The critical behaviour associated with the order-dis\-order transition can
be also observed in the staggered chirality and in its susceptibility,
defined in Eqs.~(\ref{e.chir}), (\ref{e.staggchir}),
and~(\ref{e.staggsuscchir}), and shown in Fig.~\ref{f.chiral} for
$\lambda=0$. At quite low temperatures the system displays chirality order,
witnessed by the high value of $\kappa$. As the temperature raises, the
number of cells with small chirality increases, and domains with opposite
staggered chirality develop in the lattice. This leads to a sharp drop of
the chirality and to the divergence of the chiral susceptibility at
$t_{\rm{c}}\simeq 0.41$~.

\begin{figure}
\centerline{\psfig{bbllx=14mm,bblly=70mm,bburx=192mm,bbury=205mm,%
figure=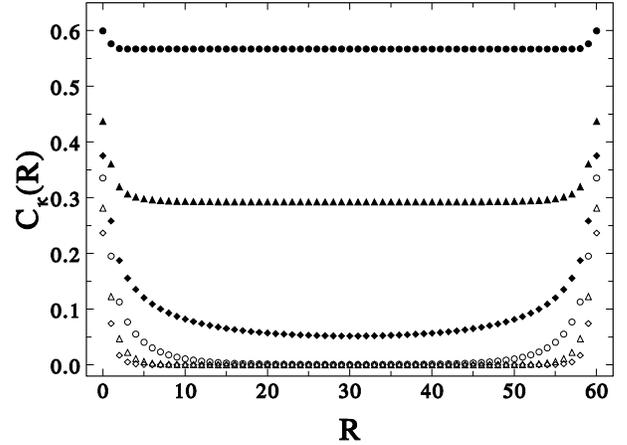,width=80mm,angle=0}}
\caption{
Chirality correlation function for $\lambda=0$, $L=60$, and various
temperatures: $t=0.3$ (full circles), $t=0.4$ (full triangles), $t=0.42$
(full diamonds), $t=0.44$ (circles), $t=0.5$ (triangles), $t=0.6$
(diamonds). }
\label{f.corrchir}
\end{figure}

The transition is also shown by the behaviour of the correlation function
of the staggered chirality, namely
\begin{equation}
C_{\bf \kappa}({\bf{R}}) = \left< (-)^{{\bf{R}}}
\kappa^{z}_{{\bf{r}}}\kappa^{z}_{{\bf{r}}+{\bf{R}}} \right>~,
\end{equation}
${\bf{R}}$ being one of the Bravais vectors of the dual lattice. This is
shown in Fig.~\ref{f.corrchir} for $\lambda=0$, $L=60$, and several
temperatures. A low temperature phase in which the system is completely
correlated is evident for $t \lesssim 0.42$. Increasing the temperature,
the correlation functions fall off exponentially. Of course it is hard to
extract from the correlation functions an accurate estimate of the critical
temperature since, approaching $t_{\rm{c}}$ from above, when the
correlation length becomes larger than the sampled lattice size, the system
behaves as if it were correlated, even if, in the thermodynamic limit, it
may be not. However, we can approximatively locate the critical region, for
$\lambda = 0$, between $t=0.4$ and $t=0.42$, consistently with the
behaviour of the specific heat, the chirality and its susceptibility.

Let us turn now to the rotational degrees of freedom. We recall that the
peculiarities of the BKT transition can be summarized as follows
\cite{BKTall,CTVmcxxz}. First of all, according to Mermin-Wagner's
theorem, the system does not show any finite magnetization in
absence of an applied magnetic field, at any $t \neq 0$ below and above the
transition temperature $t_{\rm{BKT}}$. Although the system cannot display
long-range order, below the critical temperature it is characterized by
quasi long-range order, whose macroscopic consequence is the power-law
decay of the correlation functions of the in-plane spin components,
\begin{equation}
 C({\bf n})=\left<s_{{\bf{i}}}^{x}s_{{\bf{i}}+{\bf{n}}}^{x}
 + s_{{\bf{i}}}^{y}s_{{\bf{i}}+{\bf{n}}}^{y} \right>
 \propto \frac{1}{n^{\eta}}~,
\label{e.bktdecay1}
\end{equation}
where the critical exponent $\eta$ is a function of temperature and assumes
the universal value of $\eta=1/4$ at $t_{\rm{BKT}}$. This quasi
long-range order is not destroyed by the excitation of vortex-antivortex
pairs until when, raising the temperature, the pairs unbind and the system
undergoes a transition to a disordered phase with exponentially decaying
correlation functions. The in-plane correlation length and susceptibility
both diverge exponentially for $t\rightarrow{t}_{\rm{BKT}}^{+}$
\begin{eqnarray}
 \xi & \propto & a_{\xi}~e^{b_{\xi} (t-t_{\rm{BKT}})^{-1/2}}~,
\label{e.xibkt}
\\
 \chi & \propto & a_{\chi}~e^{b_{\chi} (t-t_{\rm{BKT}})^{-1/2}}~,
\label{e.chibkt}
\end{eqnarray}
(where $\chi({\bf{K}})$ can also be taken for $\chi$) and are infinite for
$t\leq{t}_{\rm{BKT}}$. These properties have been also observed in the XY
TAF \cite{MiyashitaS84,LeeJNL8486}, which shares a similar
low-temperature phase with the ferromagnetic counterpart, if the sublattice
magnetization of the former replaces the uniform magnetization of the
latter; for example, the low-temperature phase is well described by
Eq.~(\ref{e.bktdecay1}), where ${\bf{i}}$ and ${\bf{i}}+{\bf{n}}$ belong to
the same sublattice.

Another important property of the BKT transition in unfrustrated planar
systems is the behaviour of the specific heat, which displays a maximum
slightly above the transition temperature (usually at $ t\simeq 1.1 \div
1.2~t_{\rm{BKT}}$\cite{CTVmcxxz}). However this maximum cannot be observed
in frustrated planar systems \cite{MiyashitaS84,LeeJNL8486}, where it is
hidden by the divergence connected with the chirality transition.

\begin{figure}
\centerline{\psfig{bbllx=15mm,bblly=70mm,bburx=192mm,bbury=206mm,%
figure=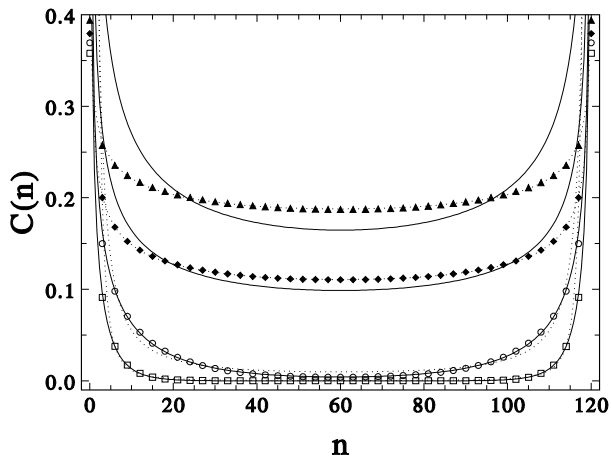,width=80mm,angle=0}}
\caption{
Spin correlation function for $\lambda=0.9$, $L=120$ and different
temperatures: $t=0.32$ (full triangles), $t=0.35$ (full diamonds), $t=0.36$
(circles), and $t=0.40$ (squares). Lines are best fits against
Eq.~(\protect\ref{e.bktdecay1}) (dashed lines) and
Eq.~(\protect\ref{e.expdecay}) (full lines). All fitting functions were
properly symmetrized to take into account the periodic boundary conditions
applied to the simulated finite-size lattice. }
\label{f.spcorfunc}
\end{figure}

\begin{figure}
\centerline{\psfig{bbllx=16mm,bblly=70mm,bburx=192mm,bbury=206mm,%
figure=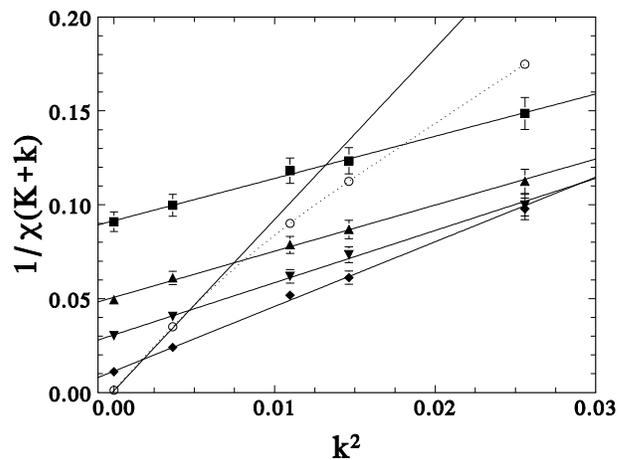,width=80mm,angle=0}}
\caption{
$\chi({\bf{K}}{+}{\bf{k}})^{-1}$ against $k^2$, for
$\lambda=0.9$ and $L=120$, at different temperatures: $t=0.35$ (open
circles), 0.36 (diamonds), 0.37 (down triangles), 0.38 (up triangles), and
0.40 (squares). The maximum value of $k^2$, for the chosen set of wave
vector shells, is $(4\pi/\protect\sqrt{3})^2\times(7/L^2)\simeq{0.0256}$,
so that a linear behaviour (full lines) against $k^2$ is expected until
$\xi^2(t)\ll{1}/0.0256\simeq{39}$. Indeed, this is not the case for
$t=0.35$, when $\xi\simeq{90}$ and the linear fit fails. The dashed line is
a guide for the eye. }
\label{f.chifit}
\end{figure}

\begin{figure}
\centerline{\psfig{bbllx=18mm,bblly=27mm,bburx=185mm,bbury=256mm,%
figure=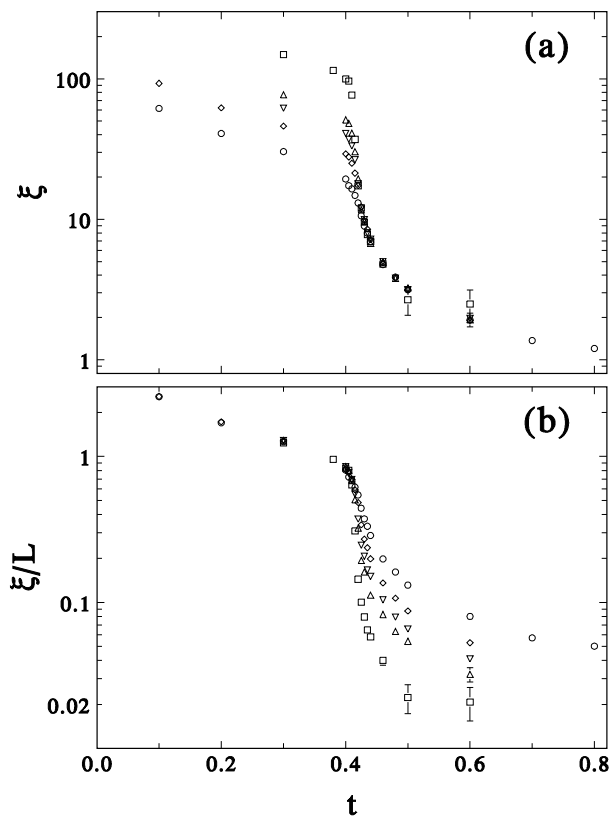,width=80mm,angle=0}}
\caption{
In-plane correlation length $\xi$ (a) for the XX0 model as a function of
temperature, for different simulation box sizes. In figure (b) $\xi/L$ is
reported to show the finite-size scaling below $t_{\rm{BKT}}$. Symbols as
in Fig.~\protect\ref{f.intenergy}. }
\label{f.corrlenght}
\end{figure}

\newpage
Fig.~\ref{f.spcorfunc} displays the correlation functions of the
in-plane spin components $C({\bf{n}})$ for $\lambda=0.9$ and various
temperatures, for $L=120$. Fittings against Eq.~(\ref{e.bktdecay1}) and
Eq.~(\ref{e.expdecay}) are also shown in the figure. The data for $t=0.32$
and $t=0.35$ can be fitted only by the power law, while for $t=0.36$ and
$t=0.4$ the exponential decay fits best: given the rather large lattice
size, it can be reasonably argued that the critical temperature
$t_{\rm{BKT}}(\lambda{=}0.9)$ is located in between between $t=0.35$ and
$t=0.36$. As already noticed, it is difficult to extract accurate
informations about $t_{\rm{BKT}}$ from the correlation functions since it
is not possible to discriminate between the high- and low-temperature
predicted behaviour unless data for lattice sizes $L>\xi$ are available,
requirement which cannot be achieved close to the critical temperature.

As already said the methods we used to extract the value of the correlation
length $\xi$ were the fit, according to Eq.~(\ref{e.chitrans}), of the total
in-plane $\bf k$-dependent susceptibility $\chi\big({\bf{k}})$, around one
of the ordering wave vectors ${\bf{K}}$, or, alternatively, the use of
Eq.~(\ref{e.xifast}). We remind that the correlation length of the infinite
system (above $t_{\rm{BKT}}$) is well defined by Eq.~(\ref{e.expdecay});
for a finite system, the same is still true for temperatures and lattice
sizes large enough that the finite-size effects are negligible; otherwise
Eq.~(\ref{e.xifast}) can be considered an {\em ad hoc} definition of $\xi$.
Provided that $\xi\lesssim{L}/6$, the values of correlation length we got
by the two methods were the same within the uncertainties. When finite-size
effects become relevant, the results differ from each other of about 4-8\%;
the first method being less reliable since Eq.~(\ref{e.chitrans}) is valid
for $\xi^{-2}k^{2}\ll{1}$, a condition which cannot be satisfied beyond the
wave vectors closest to ${\bf{K}}$, as it can be seen in
Fig.~\ref{f.chifit}: the values of $\xi$ reported in
Fig.~\ref{f.corrlenght}, for $\lambda=0$, are those obtained in the second
way. The correlation length below $t_{\rm{BKT}}$, as well as in a
neighbourhood above it, has been evaluated just to check where finite-size
effects become relevant and in order to verify the finite-size scaling law
$\xi\propto{L}$.

\begin{figure}
\centerline{\psfig{bbllx=56mm,bblly=32mm,bburx=151mm,bbury=265mm,%
figure=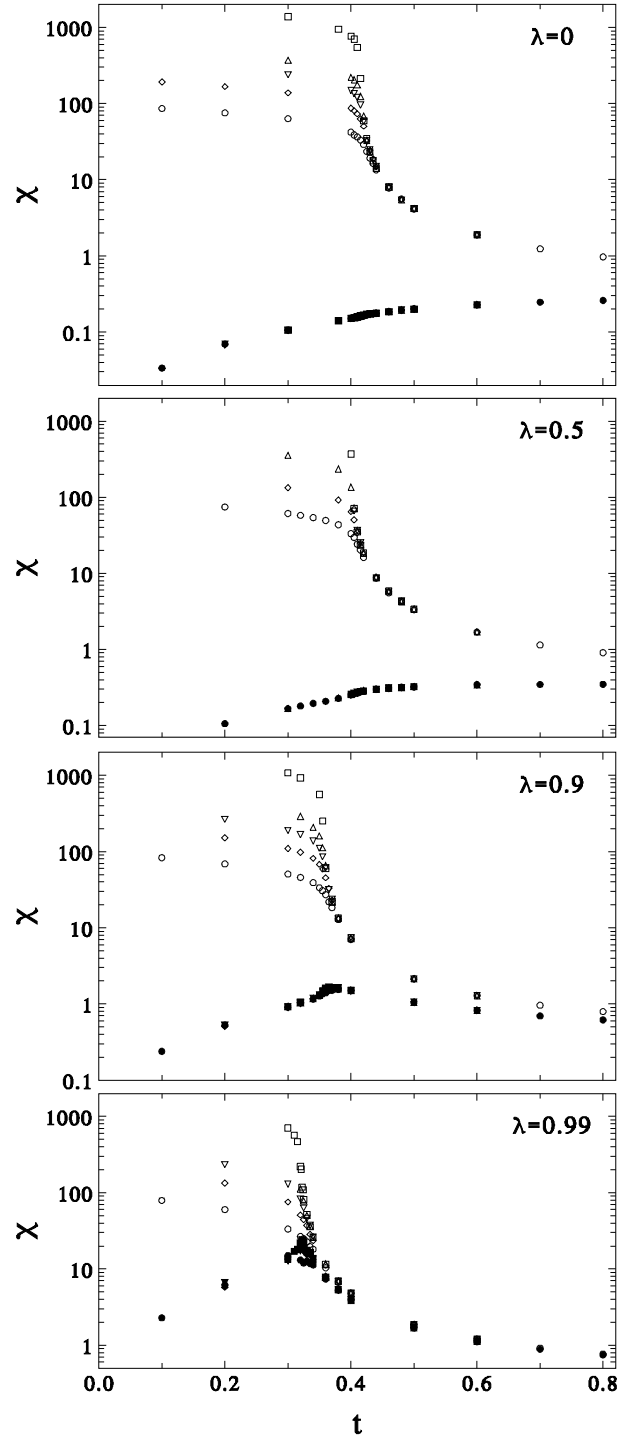,width=80mm,angle=0}}
\caption{
In-plane (open symbols) and out-of-plane (full symbols) susceptibilities
as functions of temperature, for the values of $\lambda$ reported in each
figure. Different symbols refer to the simulation box sizes as in
Fig.~\protect\ref{f.intenergy}. }
\label{f.suscept}
\end{figure}

Fig.~\ref{f.suscept} shows the in-plane ($\chi$) and out-of-plane
($\chi^{zz}$) sublattice susceptibilities, defined in Eq.~(\ref{e.chimc}),
as functions of temperature, for different values of $\lambda$ and of the
lattice sizes. As expected, the susceptibility displays finite size-effects
like the correlation length and the rule of thumb $\xi \lesssim L/6$, for
neglecting such effects also applies.
The behaviour of the out-of-plane sublattice susceptibility $\chi^{zz}$,
shown in the same figure, is also an interesting feature of the BKT
transition, also present in the XXZ model on bipartite
lattice\cite{CTVmcxxz}, which, of course, has no counterpart in the XY
models. As expected, the absolute magnitude of $\chi^{zz}$ increases, as
lambda increases, the system becoming more isotropic and the out-of-plane
fluctuations becoming easier. However, for every value of $\lambda \neq 1$,
the easy-plane character of the system prevails at low temperatures and for
$t\rightarrow0$, $\chi^{zz}\rightarrow 0$; on the other hand, in the
opposite limit, the effects of the anisotropy of the interactions
disappear, all spins can fluctuate independently from each other and both
the in-plane and out-of-plane susceptibilities approach the common value
$1/3$. Starting from the high temperature limit, and coming to lower
temperatures, $\chi$ increases whatever the value of $\lambda \neq 1$ is,
ending to diverge at $t_{\rm{BKT}}$; of course, the smaller is the
anisotropy, the longer $\chi^{zz}$ follows the behaviour of $\chi$. The
result is that, as $\lambda\rightarrow 1$, $\chi^{zz}$ develops a sharper
and sharper maximum.

\subsection{Critical temperatures}
\subsubsection{Order-disorder transition}

As we have seen in the previous sections the behaviour of the specific
heat, chirality and its susceptibility as functions of temperature
indicates the presence of an Ising-like phase transition connected with the
loss of the chirality order for every value of $\lambda < 1$ considered.
Those thermodynamic quantities allow us to estimate immediately, at least
approximately, the corresponding critical temperatures, $t_{\rm{c}}$, since
the transition appears rather sharp for all the values of $\lambda$. A
finite-size scaling analysis of these data shows that the features of such
phase-transition are consistent with two-dimensional Ising exponents.

\begin{figure}
\centerline{\psfig{bbllx=18mm,bblly=70mm,bburx=193mm,bbury=205mm,%
figure=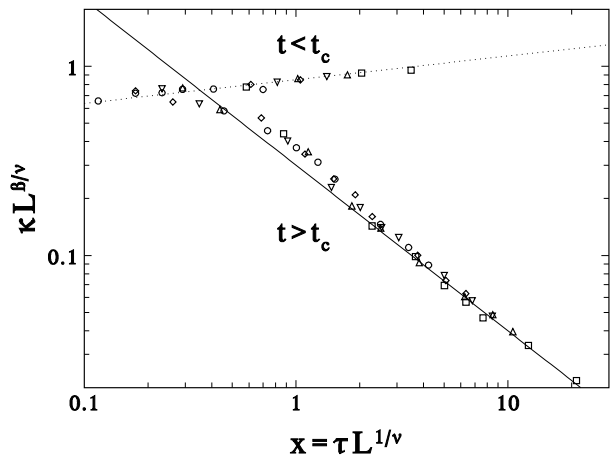,width=80mm,angle=0}}
\caption{
Finite-size scaling analysis for the staggered chirality $\kappa$, for
$\lambda=0$. Ising exponents $\beta=1/8$ and $\nu=1$ are used. Different
symbols refer to the simulation box sizes as in
Fig.~\protect\ref{f.intenergy}. }
\label{f.scalingcrl}
\end{figure}

In Fig.~\ref{f.scalingcrl} $\kappa{L}^{\beta/\nu}$ is reported, for
$\lambda=0$ and $t_{\rm{c}} = 0.412$, as a function of the reduced variable
$x=\tau L^{1/\nu}$; where $\tau$ is $1-t/t_{\rm{c}}$ for $t<t_{\rm{c}}$ and
$1-t_{\rm{c}}/t$ otherwise\cite{LeeJNL8486,Landau76}. According to the
scaling-hypothesis, close to the infinite lattice $t_{\rm{c}}$ the order
parameter $\kappa$ is given by
\begin{equation}
 \kappa = L^{-\beta/\nu}f(x)~,
\label{e.crlscal}
\end{equation}
where, since for $L\rightarrow\infty$ the power-law singularities are to be
reproduced, the limiting form of the function $f(x)$ for
$x\rightarrow\infty$ and $t<t_{\rm{c}}$ is
\begin{equation}
 f(x)\propto x^{\beta}~.
\end{equation}
Therefore, reporting as in Fig.~\ref{f.scalingcrl} $\kappa L^{\beta/\nu}$
versus $x$, the data for the various lattice sizes do collapse onto a
single curve, which is of course $f(x)$ if the values of $t_{\rm{c}}$,
$\beta$, and $\nu$ are correct. In the present case using the critical
exponents of the two-dimensional Ising model, $\nu=1$ and $\beta=1/8$, a
good agreement with the finite-size scaling law is obtained. Below
$t_{\rm{c}}$ the large-$x$ behaviour reproduces the correct critical
behaviour, which, in a double-logarithmic scale, is represented by a line
with slope $\beta=1/8$. Above $t_{\rm{c}}$ the asymptotic behaviour
shows the $1/L$ decay of the order parameter \cite{Landau76} as
$L\rightarrow\infty$ which means, according to Eq.~(\ref{e.crlscal}),
$f(x)\propto{x}^{\beta-1}$. This corresponds to the line with slope $-7/8$
in the figure. In this way it has been possible to give the estimates of
the critical temperature reported in Table~\ref{t.critemp}.

\subsubsection{BKT transition}

In order to estimate the critical temperature associated with the BKT
transition we relied on two methods: firstly, the fit of the correlation
length and the in-plane susceptibility with Eqs.~(\ref{e.xibkt}) and
(\ref{e.chibkt}), using the MC data that are representative for their
thermodynamic limit, i.e., for $t\gtrsim{t}_{\rm{BKT}}$;
secondly, we used the finite-size scaling law
\begin{equation}
 \chi \propto L^{2 - \eta(t)}~,
\label{e.chieta}
\end{equation}
which is valid for $t\lesssim{t}_{\rm{BKT}}$, to obtain the scaling
exponent $\eta(t)$, that satisfies $\eta(t_{\rm{BKT}})=1/4$. Noteworthy,
the latter method makes use of a different and independent data set.

As for the first method, we have tested for BKT behaviour, i.e., with
Eqs.~(\ref{e.xibkt}) and~(\ref{e.chibkt}), both the correlation length
$\xi(t)$ and the susceptibility $\chi(t)$. As said above, in doing this
only the reliable estimates of the thermodynamic limit of these quantities
were kept, after the criterion $\xi(t)\lesssim{6}L$, in order to discard
the data affected by finite-size effects. Of course, this has prevented us
from obtaining useful estimates $\xi(t)$ and $\chi(t)$ for $t$ close to the
transition temperature, due to the large computer-time required to simulate
large systems. Indeed, when simulating a larger lattice, besides the
increase of the time needed for any move ($\propto{N}=L^2$), we must also
face the increase of MC fluctuations and correlation time, so that much
more sample configurations should be generated in order to keep
uncertainties at a reasonable level. This would imply resorting to
large-scale simulations which is well beyond our purposes. The results for
the BKT transition temperatures from the fits of the representative data
are summarized in Table~\ref{t.tbktfit}; since $\chi(t)$ is a direct
outcome of the MC simulation, its values and the consequently fitted values
of $t_{\rm{BKT}}$ are more accurate than those for $\xi(t)$, which must
ideed be derived by fitting the MC outcomes for the ${\bf{k}}$-dependent
susceptibility. The uncertainties account both for the statistical error
and for the instability of the fit against exclusion of the data points at
the lowest temperature, where $\xi(t)\sim{L/6}$.

\begin{figure}
\centerline{\psfig{bbllx=15mm,bblly=70mm,bburx=194mm,bbury=207mm,%
figure=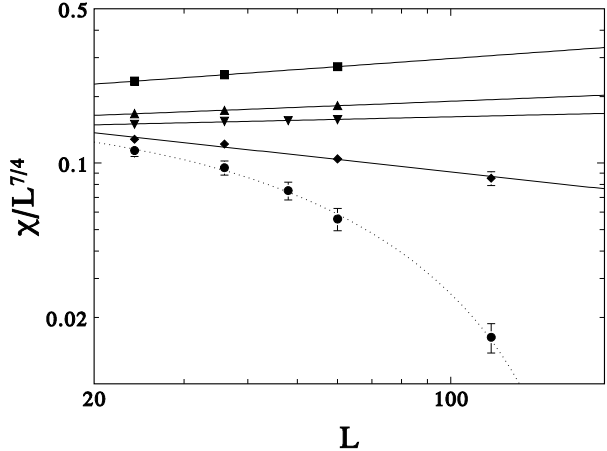,width=80mm,angle=0}}
\caption{
In-plane susceptibility over $L^{7/4}$ vs $L$, for $\lambda=0.5$, at
different temperatures around the critical one: $t=0.30$ (squares), 0.38
(up triangles), 0.39 (down triangles), 0.40 (diamonds), and 0.41
(circles). }
\label{f.chieta}
\end{figure}

As for the use of Eq.~(\ref{e.chieta}), in actual numerical calculations it
holds also slightly above $t_{\rm{BKT}}$, when $L$ is still smaller than
the thermodynamic $\xi$ and the system is already correlated. In fact this
scaling relation allows us to give an estimate of the parameter $\eta(t)$
and, by looking at which temperature such quantity attains the value 1/4,
to have an independent check of the estimated critical temperature. In
Fig.~\ref{f.chieta}, $\chi/L^{7/4}$ is plotted on a doubly logarithmic
scale as a function of the lattice size $L$ for $\lambda=0.50$. The data
fall clearly on a straight line for $t\leq{0.40}$, the slope of the lines
being the corresponding values of $1/4-\eta$; they already depart from
linearity, instead, for $t=0.41$. The values for the quantity
$1/4-\eta$ we obtain by fitting the susceptibility data with
Eq.~(\ref{e.chieta}) are reported, for the various values of $\lambda$
considered, in Table~\ref{t.tbkteta}. By interpolation, the values of
$t_{\rm{BKT}}$ appearing in the fourth column can be computed. These data
agree reasonably well with those obtained by fitting $\xi$ and $\chi$ with
Eqs.~(\ref{e.xibkt}) and Eqs.~(\ref{e.chibkt}), shown in
Table~\ref{t.tbktfit}; the trend to a slight overstimation of $t_{\rm
BKT}$ for the higher values of $\lambda$ was already observed for the
square-lattice case in Ref.~\onlinecite{CTVmcxxz}.

For every value of $\lambda$ the critical temperature $t_{\rm{c}}$ is
significantly higher than the BKT transition temperature $t_{\rm{BKT}}$,
although their difference is not much larger than the uncertainties.
Also in this respect the situation is to that of the XY TAF, where there
is no agreement in the literature on whether a single or a double
phase-transition takes place. Nevertheless, the fact that
$t_{\rm{c}}\gtrsim{t}_{\rm{BKT}}$ for all values of $\lambda$ in a
systematic way, makes it unlikely that the difference could just be due to
statistical errors.

\section{Concluding remarks}
\label{s.concl}

\begin{figure}
\centerline{\psfig{bbllx=15mm,bblly=70mm,bburx=194mm,bbury=207mm,%
figure=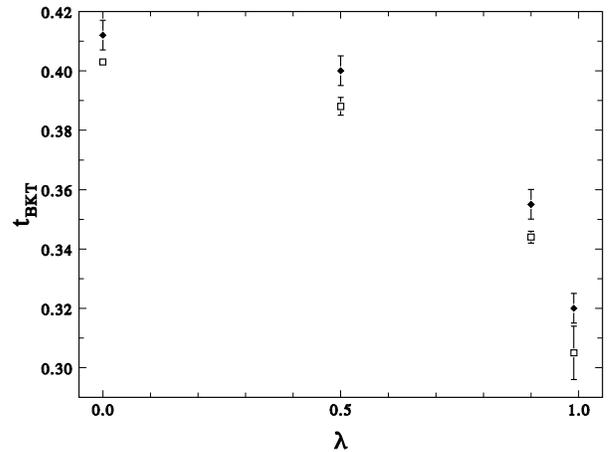,width=80mm,angle=0}}
\caption{
Transition temperatures for the Ising-like transition (full diamonds)
and for the BKT one (squares), the latter estimated through the BKT fit of
the susceptibility, vs. the anisotropy. }
\label{f.tcrit}
\end{figure}

We have performed Monte Carlo simulations of the two-dimensional XXZ model
on a triangular lattice at different values of the easy-plane anisotropy
constant $\lambda$. For every value of $\lambda$ considered, the situation
appears quite similar to that observed in the XY triangular
antiferromagnet, where frustration induces an order-disorder transition,
associated with the two-fold additional degeneracy of the ground state, and
a BKT transition connected with the sublattice in-plane orientational
ordering. The critical behaviour turns out to be consistent with an Ising
transition, for the internal energy, specific heat, chirality and the
associated susceptibility, while it is consistent with a BKT transition
with respect to the in-plane correlation length and susceptibility. The
value of both the critical temperatures decreases with anisotropy strength,
as shown in Fig.~\ref{f.tcrit}; this is consistent with the fact that the
critical behaviour observed is connected with the planar character of the
system, and that both the chirality and the orientational quasi-ordering
are disturbed by the out-of-plane fluctuations of the spins which, at a
fixed temperature, increase with the value of $\lambda$. For the same
reason the phase transition in the XX0 model takes place at a temperature
($t\simeq{0.403}$) which is sensibly lower than that observed in the XY
model ($t\simeq{0.505}$), where the spins are confined in the $xy$ plane.
As for the question of whether a single or two phase transitions are
occurring, our results for the transition temperatures (Fig.~\ref{f.tcrit})
support the second hypotesis, consistently also with the most recent
high-precision MC simulations of the fully frustrated XY model
\cite{Olsson9597}, where the existence of a new universality class is ruled
out.

\newpage

\begin{table}
\caption{
Critical temperature $t_{\rm{c}}$ associated to the chirality
phase-transition. }
\begin{tabular}{cdcc}
 & $\lambda$ & $t_{\rm{c}}$ & \\
\tableline
 & 0.00 & 0.412 $\pm$ 0.005 & \\
 & 0.50 & 0.400 $\pm$ 0.005 & \\
 & 0.90 & 0.355 $\pm$ 0.005 & \\
 & 0.99 & 0.320 $\pm$ 0.005 & \\
\end{tabular}
\label{t.critemp}
\end{table}

\begin{table}
\caption{
BKT transition temperature as obtained fitting the in-plane correlation
length $\xi(t)$ and the in-plane susceptibility $\chi$ against
Eqs.~(\protect\ref{e.xibkt}) and~(\protect\ref{e.chibkt}). }
\begin{tabular}{ccccc}
 & $\lambda$ & $t_{\rm{BKT}}$ ($\xi$ fit) & $t_{\rm{BKT}}$ ($\chi$ fit) & \\
\tableline
 & 0.00 &  0.402 $\pm$ 0.002 &  0.403 $\pm$ 0.001 & \\
 & 0.50 &  0.391 $\pm$ 0.002 &  0.388 $\pm$ 0.003 & \\
 & 0.90 &  0.345 $\pm$ 0.006 &  0.344 $\pm$ 0.002 & \\
 & 0.99 &  0.306 $\pm$ 0.008 &  0.305 $\pm$ 0.009 & \\
\end{tabular}
\label{t.tbktfit}
\end{table}

\begin{table}
\caption{
Scaling exponent $\eta(t)$ as obtained by direct finite-size scaling
analysis of the in-plane susceptibility data according to
Eq.~(\protect\ref{e.chieta}). The values of $t_{\rm{BKT}}$ given in the
fourth column are obtained by interpolation of the function $\eta(t)$ at
the point $\eta=1/4$. }
\begin{tabular}{dcd@{\,$\pm$~}lc}
$\lambda$ & $t$  & \multicolumn{2}{c}{1/4 - $\eta$} & $t_{\rm{BKT}}$ \\
\tableline
 0.00 & 0.300  &  0.165  & 0.002  &                   \\
      & 0.400  &  0.054  & 0.005  &                   \\
      & 0.405  &  0.051  & 0.007  &                   \\
      & 0.410  & -0.06   & 0.01   & 0.407 $\pm$ 0.003 \\
 0.50 & 0.300  &  0.1654 & 0.0007 &                   \\
      & 0.380  &  0.09   & 0.04   &                   \\
      & 0.390  &  0.052  & 0.008  &                   \\
      & 0.400  & -0.25   & 0.03   & 0.391 $\pm$ 0.005 \\
 0.90 & 0.300  &  0.1486 & 0.0005 &                   \\
      & 0.320  &  0.13   & 0.06   &                   \\
      & 0.340  &  0.068  & 0.008  &                   \\
      & 0.350  &  0.00   & 0.01   & 0.350 $\pm$ 0.005 \\
 0.99 & 0.200  &  0.204  & 0.009  &                   \\
      & 0.300  &  0.131  & 0.02   &                   \\
      & 0.310  &  0.08   & 0.01   &                   \\
      & 0.315  & -0.02   & 0.01   & 0.314 $\pm$ 0.002 \\
\end{tabular}
\label{t.tbkteta}
\end{table}


\newpage

\begin{thebibliography}{200}

\bibitem[*]{e-LC} Present address: Scuola Internazionale Superiore
di Studi Avanzati, Via Beirut 2-4, 34013 Trieste, Italy.
Electronic address:~ capriotti@fi.infn.it, caprio@sissa.it
\bibitem[\dagger]{e-RV} Electronic address:~ vaia@ieq.fi.cnr.it
\bibitem[\ddagger]{e-AC} Electronic address:~ cuccoli@fi.infn.it
\bibitem[\S]{e-VT} Electronic address:~ tognetti@fi.infn.it

\bibitem{KawamuraM84}
H. Kawamura and S. Miyashita, J. Phys. Soc. Jpn. {\bf 53}, 4138 (1984).

\bibitem{AzariaDM92}
P. Azaria, B. Delanotte, and D. Mouhanna, Phys. Rev. Lett. {\bf 68}, 1762
 (1992).

\bibitem{WintelEA95}
M. Wintel, H. U. Everts, and W. Apel, Phys. Rev. B {\bf 52}, 13480 (1995).

\bibitem{MiyashitaS84}
S. Miyashita and H. Shiba, J. Phys. Soc. Jpn. {\bf 53}, 1145 (1984).

\bibitem{LeeJNL8486}
D. H. Lee, J. D. Joannnopoulos, J. W. Negele, and D. P. Landau,
Phys. Rev. Lett. {\bf 52}, 433 (1984); Phys. Rev. B {\bf 33}, 450 (1986).

\bibitem{TeitelJ83}
S. Teitel and C. Jayaprakash, Phys. Rev. B {\bf 27}, 598 (1983).

\bibitem{Olsson9597}
P. Olsson, Phys. Rev. Lett. {\bf 75}, 2758 (1995) and {\bf 77}, 4850 (1997);
Phys. Rev. B {\bf 55}, 3585 (1997).

\bibitem{CGTVV95}
A. Cuccoli, R. Giachetti, V. Tognetti, P. Verrucchi, and R. Vaia,
 J. Phys.: Condens. Matter {\bf 7}, 7891 (1995).

\bibitem{BCTVV96}
C. Biagini, A. Cuccoli, V. Tognetti, P. Verrucchi, R. Vaia,
 J. Appl. Phys. {\bf 79}, 4638 (1996).

\bibitem{CTVV96prl}
A. Cuccoli, V. Tognetti, P. Verrucchi, and R. Vaia,
 Phys. Rev. Lett. {\bf 77}, 3439 (1996), and {\bf 79}, 1584 (1997).

\bibitem{CCTVV97}
L. Capriotti, A. Cuccoli, V. Tognetti, R. Vaia, and P. Verrucchi,
Physica D (in press, 1997).

\bibitem{MerminW66}
N. D. Mermin and H. Wagner, Phys. Rev. Lett. {\bf 17}, 1133 (1966).

\bibitem{BKTall}
V. L. Berezinskii, Zh. Eksp. Teor. Fiz. {\bf 59}, 907 (1970);
J. M. Kosterlitz and D. J. Thouless, J. Phys. C {\bf 6}, 1181 (1973);
J. M. Kosterlitz, J. Phys. C {\bf 7}, 1046 (1974).

\bibitem{Wannier50}
G. H. Wannier, Phys. Rev. {\bf 79}, 357 (1950).

\bibitem{PelcovitsN76}
R. A. Pelcovits and D. R. Nelson, Phys. Lett. {\bf 57 A}, 23 (1976)

\bibitem{Khokhlacev76}
S. B. Khokhlacev, Zh. Eksp. Theor. Fiz. {\bf 70}, 265 (1976).

\bibitem{HikamiT80}
S. Hikami and T. Tsuneto, Prog. Theor. Phys. {\bf 63}, 387 (1980).

\bibitem{CTVmcxxz}
A. Cuccoli and V. Tognetti, R. Vaia, Phys. Rev. B {\bf 52}, 10221 (1995).

\bibitem{Wysin94}
G. M. Wysin, Phys. Rev. B {\bf 49}, 8780 (1994).

\bibitem{Metropolis53}
N. Metropolis {\em et al.} J. Chem. Phys. {\bf 21} 1087 (1953).

\bibitem{BrownC87}
F. R. Brown and T. J. Woch, Phys. Rev. Lett. {\bf 58}, 2394 (1987).

\bibitem{Creutz87}
M. Creutz, Phys. Rev. D {\bf 36}, 515 (1987).

\bibitem{MadrasS88}
N. Madras and A. D. Sokal, J. Stat. Phys. {\bf 50},109 (1988).

\bibitem{Barber83}
M. N. Barber, in {\em Phase Transitions and Critical Phenomena}, edited
by C. Domb and J. L. Lebowitz (Academic, London, 1983), Vol. 8, pp.
145-266.

\bibitem{Landau76}
D. P. Landau, Phys. Rev. B, {\bf13}, 2997 (1976).



\end{thebibliography}
\end{document}